\documentclass[twocolumn]{jpsj2}

\def\runtitle{%
Electronic Structure of transition metals Fe, Ni and Cu
in the $GW$ approximation
}
\def\runauthor{%
Atsushi \textsc{Yamasaki}
and 
Takeo \textsc{Fujiwara}
}

\title{%
Electronic Structure of transition metals Fe, Ni and Cu
in the $GW$ approximation
}

\author{%
Atsushi \textsc{Yamasaki}
\thanks{Present address: Max-Planck-Institut f\"{u}r Festk\"{o}rperforschung,
        D-70569 Stuttgart, Germany.
        E-mail: A.Yamasaki@fkf.mpg.de}
Takeo \textsc{Fujiwara}
\thanks{E-mail address: fujiwara@coral.t.u-tokyo.ac.jp}
}

\inst{%
Department of Applied Physics, University of Tokyo,   
Bunkyo-ku, Tokyo 113-8656
}

\recdate{October 4, 2002}

\abst{%
The quasiparticle band structures of 3$d$ transition metals,
ferromagnetic Fe, Ni and paramagnetic Cu, are calculated by the $GW$
approximation. 
The width of occupied 3$d$ valence band, which is overestimated in the
LSDA, is in good agreement with experimental observation.
However the exchange splitting and satellite in spectra are not
reproduced and it is required to go beyond the $GW$ approximation.
The effects of static screening and dynamical correlation are discussed
in detail in comparison with the results of the static COHSEX
approximation.
The dynamical screening effects are important for band width narrowing.
}

\kword{%
Density functional theory, GW approximation, LMTO, product-basis,
transition metal, dynamical screening
}

\begin{document}   
\sloppy
\maketitle

%%%%%%%%%%%%% Section1 %%%%%%%%%%%%%%%%%%%%%%%%%%%%%%%%%%%%%%%%
\section{Introduction} \label{sec1:intro}

Lattice structure, lattice constants and bulk
moduli in 3$d$ transition metal are well described by the
local-spin-density approximation (LSDA)~\cite{LSDA1,LSDA2}
or the generalized gradient approximation (GGA)~\cite{GGA1,GGA2}.
However, the occupied 3$d$ band width is too broad, and the exchange
splitting is overestimated.
The good agreement is essentially related with the property of the
ground-state, and the discrepancies are associated with excitation
properties. 

%---------------------
The $GW$ approximation (GWA) is based on the many-body perturbation
theory~\cite{GW1,GW2,GW3} and can describe the quasiparticle property.
The self-energy of GWA is the first term in a series expansion of
dynamical correlation and it is treated by the random-phase
approximation (RPA). 

The plane wave basis set based on the pseudopotential method is used in
many $GW$ calculations.
In simple metals and semiconductors,
the single plasmon peak is often assumed within the plane wave
framework (the plasmon pole approximation).~\cite{GW-semicon1}
However the plasmon peak of transition metal cannot be well-defined
isolated peak due to interband transition in the same energy region.
The transition metal has strong atomic potential for 3$d$
electrons, the 3$d$ orbital is localized and 
the plane wave formalism cannot be applied. 
Moreover it is essentially important to include core electrons in many
cases.  
Therefore the plasmon peak approximation is not applicable 
to the dielectric function of transition metals and all-electron
calculation and localized orbital basis set are needed.

%---------------------------
In this paper, the $GW$ method based on the linear muffin-tin orbital 
(LMTO) method~\cite{OKA} and the product-basis method~\cite{pd-basis} 
are applied to the series of transition metal.
There is a numerical difficulty in the $\bf k$-point summation of
self-energy with the momentum transfer ${\bf q}\cong 0$.
This summation is treated by the offset method,~\cite{offset} 
and test calculation of the exchange energy in the electron gas is
performed.
The paper is organized as follows. 
The theoretical framework is described in \S~\ref{sec2:thory}.
The numerical technique and test calculation of electron gas are also
given in this section.
The results for these systems and detailed discussion are presented in
\S~\ref{sec3:result}. 
Finally, in \S~\ref{sec4:summary} we present our summary.

%%%%%%%%%%%%% Section2  %%%%%%%%%%%%%%%%%%%%%%%%%%%%%%%%%%%%%%%%
\section{Theory} \label{sec2:thory}

%--------- GW approximation  -----------------------------------
\subsection{$GW$ approximation}

In the GWA the self-energy is replaced by the lowest order term of 
the expansion as $\Sigma(1,2) = i G(1,2) W(1,2)$.
$G$ is the one particle Green function and
the dynamically screened interaction $W$ is defined by
\begin{align}
 W(1,2) &= \int d(3) \epsilon^{-1}(1,3) v(3,2) \label{dyn-int}\\
 &= v(1,2) + \int d(34) v(1,3) \chi^0(3,4) W(4,2), \label{dyn-int2}
\end{align}
where $\epsilon^{-1}$ is the inverse dynamical dielectric function, $v$
is the bare Coulomb potential and $\chi^0$ is the irreducible
polarization function $\chi^0(1,2) = -i G(1,2)G(2,1)$.
Here we use an abbreviated notation 
$(1)=({\bf r}_1,\sigma_1, t_1)$ and
$v(1,2)=v({\bf r}_1,{\bf r}_2) \delta(t_1-t_2)$.
Equation (\ref{dyn-int2}) is treated by the RPA.

We adopt the LSDA Hamiltonian to be the unperturbed one
$H^0=T+V^H+V^{xc}_{LSDA}$.
Here $T$ is the kinetic energy, $V^H$ is the Hartree potential,
and $V^{xc}_{LSDA}$ is the exchange-correlation potential in the LSDA.
We presume the wavefunctions $\{ \psi_{{\bf k}n}({\bf r})\}$ 
of the LSDA to be a reasonably good starting wavefunctions.
Then the self-energy can be written by three terms as
$\Delta\Sigma = \Sigma^x + \Sigma^c - V^{xc}_{LSDA},$
where $\Sigma^x (= i G v)$ is the exchange part (the Fock term) and 
$\Sigma^c (= i G W^c)$ is the dynamical correlation part. 
$W^c$ is the second term in eq.(\ref{dyn-int2}).
The quasiparticle energy is given as
\begin{equation}
 E_{{\bf k}n} = \epsilon_{{\bf k}n} + Z_{{\bf k}n} 
  \Delta\Sigma_{{\bf k}n} (\epsilon_{{\bf k}n}) ,
\end{equation}
where $\epsilon_{{\bf k}n}$ is the LSDA eigenvalue. 
The self-energy is 
$\Delta\Sigma_{{\bf k}n} (\epsilon_{{\bf k}n}) = 
 \langle \psi_{{\bf k}n} | \Sigma^x + \Sigma^c(\epsilon_{{\bf k}n}) -
 V^{xc}_{LSDA} | \psi_{{\bf k}n} \rangle$
and the renormalization factor is 
$Z_{{\bf k}n}= ( 1 - \partial \Delta\Sigma_{{\bf k}n}(\epsilon_{{\bf k}n})
/ \partial\omega )^{-1}$. 
The renormalization factor $Z_{{\bf k}n}$ is a measure of 
the occupation number and should equal to the discontinuity of
occupation number at the Fermi energy. 
Therefore it should satisfy the condition $Z_{{\bf k}n}\leq 1$.
In the present work we perform one iteration calculation without
self-consistency.~\cite{GW-SCF}
%-

%--------- LMTO minimal basis set and Choice of LMTO parameters}--------
\subsection{LMTO minimal basis set and product-basis}\label{sec2-b:LMTO}

Because the plane wave basis becomes very costly 
for systems containing 3$d$ electrons, 
the LMTO method~\cite{OKA} is more appropriate.
We use the LMTO basis set $\chi_{{\bf R}L\nu}({\bf r})$ within the
atomic sphere approximation (ASA) for the LSDA calculation.
Here $L$ is angler momentum $L$ = $(l,~m)$.
The LMTO can be expanded by the muffin-tin orbital 
$\phi_{{\bf R}L\nu} ({\bf r})$ and it's energy derivative 
$\dot \phi_{{\bf R}L\nu} ({\bf r})$.

The functional space of basis for $\Sigma$ is spanned as 
\begin{equation}
 \{ \Sigma \}=\{ \psi \psi \}=\{ \chi \chi \}
  = \{ \phi \phi \}+\{ \phi {\dot \phi} \}+
  \{ {\dot \phi} {\dot \phi} \}    \ .
\end{equation}
In fact the mixing coefficients of $\dot{\phi}$ to $\phi$ are less than
$0.1$ for the most part and the norm of $\dot{\phi}$, 
$\langle \dot{\phi}\dot{\phi} \rangle$, is $0.1\sim 0.3$ even in the
largest case,  
so the terms including $\dot \phi$ can be dropped out.
More detailed description is shown in ref.~\ref{GWnonmag}.

% Figure 1----------------------------------------------------------
\begin{figure}[t]
\begin{center}
 \includegraphics[width=72mm,clip]{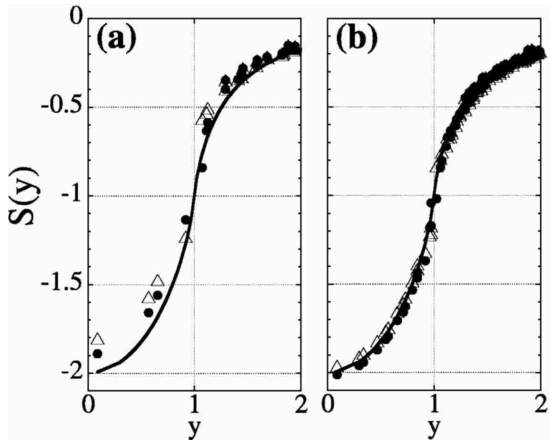}
\end{center}
\caption{Exchange energy $S(y)$ of the electron gas as a function of 
         $y=k/k_F$. 
         The solid lines for the exact result, the open triangles for
         the simple summation and the closed circles for the offset
         method. 
         The numbers of the mesh points in the Brillouin zone are 
         (a) $64=(4\times 4\times 4)$ and
         (b) $512=(8\times 8\times 8)$.}\label{EG-fig}
\end{figure}%
%-----------

\subsection{Numerical technique}

The Coulomb matrix $v({\bf q})$ has  a singularity  at ${\bf q}=0$  as 
$F({\bf q})=1/|{\bf q}|^2$.  
The integration of $v({\bf q})$ over the Brillouin zone does not 
diverge but special cares are needed not only for the ${\bf q}=0$ term
but for small finite ${\bf q}$.
For a choice of the discrete points near ${\bf q}=0$,  
we use the offset $\Gamma$-point method,~\cite{offset}
where the integration of $F({\bf q})$ over the Brillouin zone can be
performed analytically and  
the offsetted points
$\bf Q$'s are chosen near ${\bf q}=0$ so as to satisfy a relation 
\begin{equation}
 \int_{B.Z.} F({\bf q}) d{\bf q} = \sum_{\bf Q} F({\bf Q}) + 
  \sum_{{\bf k}\neq 0} F({\bf k}) .
\end{equation}
Here $\bf k$'s are the discrete mesh points in the Brillouin zone.

%------------------
The exchange energy of the electron gas system is given 
as a function of a wave vector $k$ as 
$\Sigma^x(k) =\frac{e^2 k_F}{\pi} S(y)$,
where 
$S(y) = - \left( 1+\frac{1-y^2}{2y} \ln \left| \frac{1+y}{1-y} \right|
 \right)$, $y=\frac{k}{k_F}$ and
$k_F$ is the Fermi wave vector. 
The empty lattice calculation is done with $spdf$ orbitals in the LMTO 
method. 
We calculate the exchange energy of the electron gas 
in a fcc lattice with a lattice constant $a=6.824a_0$ 
which corresponds to the fcc copper and $a_0$ is the Bohr radius. 
The corresponding electron gas parameter is $r_s=2.6668$. 
The calculated $S(y)$, by  the simple summation, by  
the offset method  and by the exact $S(y)$, are shown 
in Fig.~\ref{EG-fig}. 
In the simple summation, the diverging term
$1/|{\bf q}|^2$ is simply averaged inside a sphere of a volume
equal to that of one $\bf k$-mesh point.
The number of $\bf k$-mesh of the Brillouin-zone in the 
calculation is 
(a) 64  points ($4\times4\times4$) and
(b) 512 points ($8\times8\times8$). 
In case of 512 points of these structure, eight offsetted points 
${\bf Q}$ are $\frac{2\pi}{a}(\pm0.038,\pm0.038,\pm0.038)$.
The derivative of $S(y)$ has 
a logarithmic singularity at the Fermi energy ($y=1$).
Unphysical gap still remains at the Fermi energy in the simple
summation of these examples. 
A large number of  $\bf k$ points is necessary  for a convergence 
in the simple summation.  
But the offset method can reduce the number of $\bf k$ point for rapid
convergence even in case of small number of mesh points.
The careful treatment of the Coulomb matrix at or near ${\bf q}=0$ 
is very crucial  near the band gap or the Fermi energy.
%------------------

%-------------
% Tables 
\begin{table}[b]
\begin{center}
\caption{The Magnetic moment $\mu_{spin}(\mu_B)$ of Fe and Ni,
         the exchange splitting $\delta E_{ex}$ (eV) of Fe and Ni,
         the band width of the occupied 3$d$ valence bands
         $W_{d,occ}$ (eV) of Fe, Ni and Cu.}
\begin{tabular}{llccccc}
\hline
&     &               & LSDA & COHSEX & GW   & expt. \\
\hline
$\mu_{spin}$
&  Fe &               & 2.27 & 2.04   & 2.31 & 2.13~\cite{TM-MM1,TM-MM2} \\
&  Ni &               & 0.54 & 0.62   & 0.55 & 0.57~\cite{TM-MM1,TM-MM2} \\
\hline
$\delta E_{ex}$
&  Fe & $\Gamma_{25}$ & 2.0  & 1.7    & 1.9  & 2.1~\cite{TM-ARPES1} \\ 
&     & $H_{25}$      & 2.3  & 2.1    & 1.8  & 1.8~\cite{TM-ARPES2} \\
&     & $P_4$         & 2.2  & 2.5    & 2.3  & 1.5~\cite{TM-ARPES3} \\
&  Ni & $L_3$         & 0.6  & 0.5    & 0.7  & 0.3~\cite{TM-ARPES3} \\
&     & $X_2$         & 0.6  & 0.5    & 0.7  & 0.2~\cite{TM-ARPES4} \\
\hline
$W_{d,occ}$
&  Fe &               & 3.7  & 4.6    & 3.4  & 3.3~\cite{TM-XPS-Fe1,
                                              TM-XPS-Fe2,TM-XPS-Fe3} \\
&  Ni &               & 4.5  & 5.1    & 3.3  & 3.2~\cite{TM-XPS-Ni} \\
&  Cu &               & 3.3  & 3.7    & 2.9  & 3.0~\cite{TM-Cu} \\
\hline
\end{tabular}
\label{tbl}
\end{center}
\end{table}
%-------------

% Figure ----------------------------------------------------------
\begin{fullfigure}[t]
\begin{center}
 \includegraphics[width=165mm,clip]{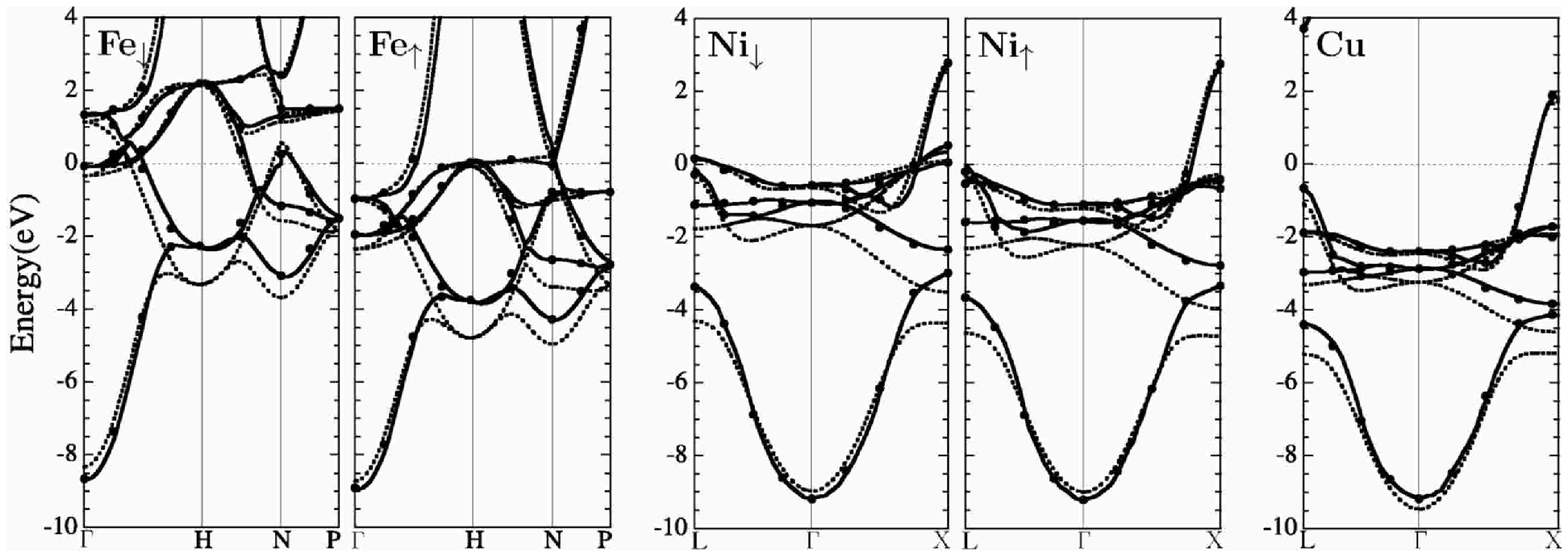}
\end{center}
\caption{The energy bands of Fe, Ni and Cu, 
         calculated by the LSDA (dotted lines) 
         and the GWA  (solid lines) along high symmetric lines. 
         The closed circles are the calculated points in the GWA.
         In Fe and Ni, left side and right side show minority spin and
         majority spin, respectively.
         The high symmetric points are 
         $\Gamma=(0,0,0)$, H$=(1,0,0)$, N$=(1/2,1/2,0)$ and 
         P$=(1/2,1/2,1/2)$ in bcc lattice (Fe), and
         L$=(1/2,1/2,1/2)$,
         $\Gamma=(0,0,0)$ and X$=(1,0,0)$ in fcc lattice (Ni and Cu).
         Fermi energy is set to zero. ($E_F$=0)}
\label{band}
\end{fullfigure}%
%-----------

%%%%%%%%%%%%% Section3  %%%%%%%%%%%%%%%%%%%%%%%%%%%%%%%%%%%%%%%%
\section{Results and Discussions} \label{sec3:result}

%--- background ---
In the calculation of LSDA, the lattice structure and constants
of Fe, Ni and Cu are bcc and $a=$ 2.87~\AA, fcc and 3.52~\AA, fcc and
3.61~\AA, respectively.~\cite{wyckoff}
The band structures of Fe, Ni and Cu, calculated both in the LSDA
and the GWA, are shown in Fig.~\ref{band} along high symmetric
lines. 
The localized 3$d$ orbital has a weak hybridization with the extended
4$s$, 4$p$ orbitals and is below Fermi energy. 

\begin{figure}[t]
\begin{center} 
 \includegraphics[width=72mm,clip]{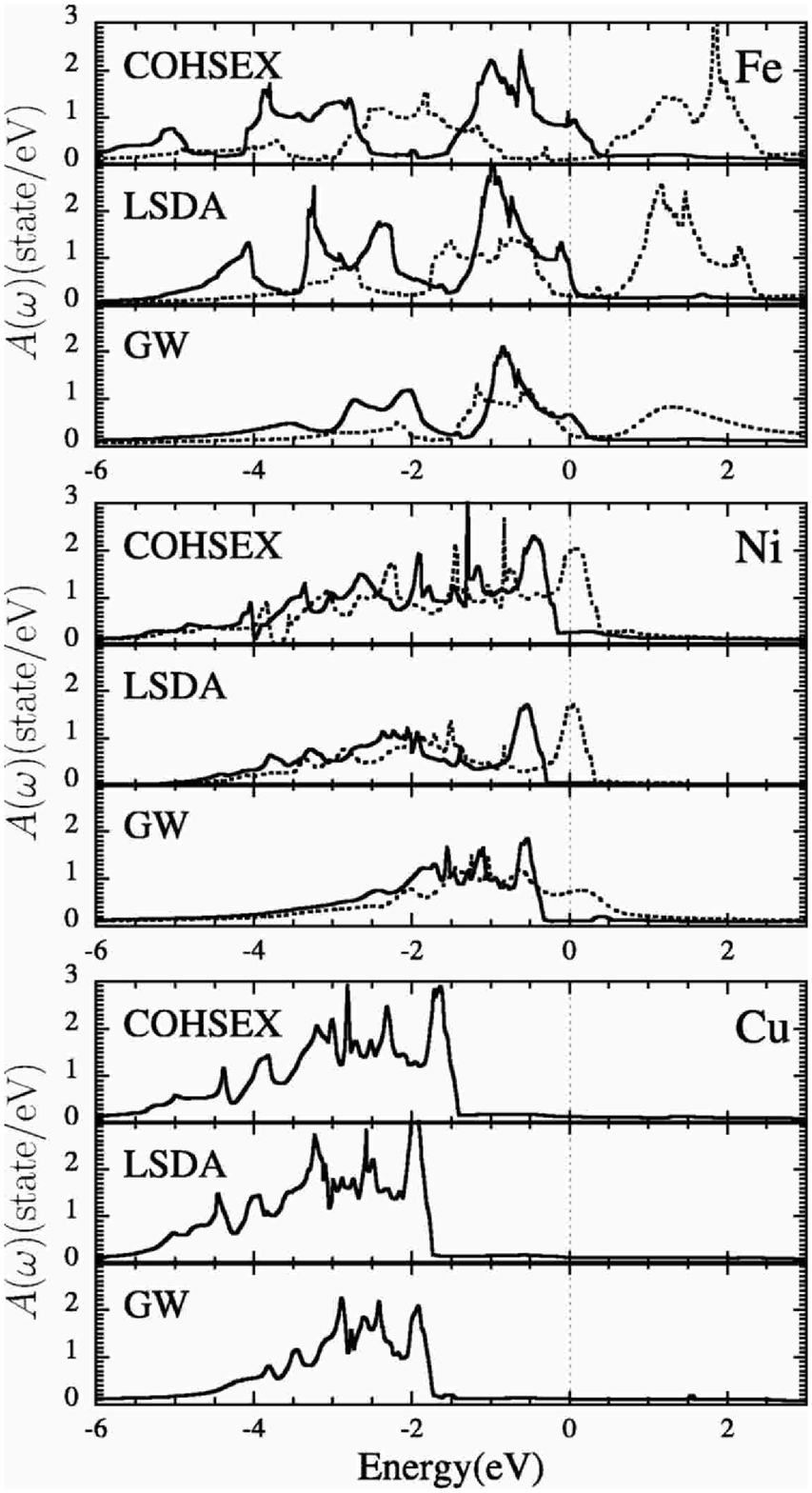}
\end{center}
\caption{The spectral function $A(\omega)$ in the transition metals Fe,
         Ni and Cu by the GWA (bottom), the LSDA (middle) and the static
         COHSEX approximation (top).
         In Fe and Ni, the solid lines show majority spin and 
         the dotted lines show minority spin, respectively.}
\label{spectra}
\end{figure}

The magnetic moment, the exchange splitting and the band width of the
occupied 3$d$ valence bands in the GWA are summarized in
Table~\ref{tbl}, in comparison with those by the LSDA and the static
COHSEX approximation.~\cite{GW2}
Our results of Ni are in good agreement with those of the previous 
$GW$ calculation.~\cite{GW-TM}
The spectral function $A(\omega)=-(1/\pi){\rm Im \, Tr} G(\omega)$ is
shown in Fig.~\ref{spectra}.

%--- band width ---
%--- dynamical effects (static COHSEX) ---
% $W_{d,occ}$
The occupied 3$d$ valence band width of transition metals Fe, Ni and Cu
in the LSDA is overestimated in comparison with experimental
observation, especially in Ni.
The valence band width is in reasonably agreement with experiment in the
GWA.
In Fig.~\ref{spectra}, the band narrowing occurs in the occupied valence
band of both the majority and the minority spin in Fe and Ni.
But the width of unoccupied 3$d$ band of Fe is unchanged in the GWA.
The source of band narrowing is the screening for the valence
electrons. 
In the spectral function of the GWA, the plasmon-like excitation appears
around 30 eV above and below the Fermi energy and also the long tail
extends over wide lower energy region.
The intensity of spectrum is totally suppressed by the excitations in
wide energy region.
The intensity of the $GW$ spectrum is actually reduced by a factor of
$Z_{{\bf k}n}$ and, in Fig.~\ref{spectra}, the reduction factor is 
$Z_d\approx0.5$--0.6.
In the Hartree Fock (HF) approximation which includes no
screening effects, the band width is overestimated.
No screening in the HF gives zero density of states at the Fermi level
in the electron gas and also gives overestimated band gap in
insulators and semiconductors.
In the static COHSEX approximation, which includes static screening,
the band width is much smaller than the HF results,
and is almost the
same as the one in the LSDA or still wider.
Moreover the 4$s$ state is located too much deep because it exists far
from Fermi level. 
The band width in the GWA is in good agreement with experiment.
We can see that the dynamical correlation effect is important for the
band width in the 
transition metals from the comparison between the GWA and the static
COHSEX approximation. 

%--- exchange splittimg and satellite ---
% discrepancy and its reason (low density limit in Ni)
The magnetic moment $\mu_{spin}$ of Fe and Ni is almost the same
as the result of LSDA and is in good agreement with experiment.
The difference of the exchange splitting $\delta E_{ex}$ between the LSDA
and experiments in Fe ($\sim30\%$) is smaller than that in Ni
($\sim50\%$). 
$\delta E_{ex}$ of Fe becomes close to the experimental value in the
GWA.
In Ni, the discrepancy of $\delta E_{ex}$ is not improved by the
GWA.
In the HF, $\delta E_{ex}$ is overestimated.
The screening effects of correlation term
$\Sigma^c$ in the GWA or the static COHSEX approximation reduce 
$\delta E_{ex}$ of HF. 
However the GWA only includes long-range correlation effects, and cannot 
describe short-range effects such as electron-electron or hole-hole
scattering process.
Higher order diagrams (e.g.~vertex corrections) is needed 
for electron-electron and hole-hole scattering.
Especially two-hole bound states are very important to the exchange
splitting and the satellite structure of spectrum if on-site Coulomb
interaction between $d$ electrons is large.~\cite{NiTmat1,NiTmat2}
The effective Coulomb interaction is obtained from an analysis of Auger
spectra, Ni is $U\approx4.0$~eV and Fe is $U\approx1.0$~eV.~\cite{Uexp}
The discrepancy between experiments and the GWA in Ni is caused by the
short-range correlation effects from large $U$.
The clear satellite in Ni can be also explained, and the GWA cannot
reproduce it. 
Since the 3$d$ band is full in Cu, there is no hole-hole correlation
and the GWA can work quite well.

%--- W(0) ---
%--- self-energy ---
% Z
The renormalization factor of transition metal 3$d$ states is 
$Z$ = $0.52$--$0.58$ in Fe, $Z$ = $0.48$--$0.53$ in Ni and $Z$ =
$0.53$--$0.66$ in Cu. 
$Z$ of 4$s$ states is about $0.7$--$0.8$ in these systems. 
Those results of the renormalization factor show that 
the interaction between 3$d$ electrons is large,
and the correlation in Ni is strongest, which is consistent with the
large Coulomb interaction $U$.

% W(0) correlation strength 
The static screened $d$-$d$ Coulomb interaction
%$\langle \phi_d \phi_d | W(\omega=0)|\phi_d \phi_d \rangle$ 
$\langle \phi_d \phi_d | $ $W(\omega=0)|\phi_d \phi_d \rangle$ % notice!
is calculated to be about 1.4~eV, 1.2~eV and 3.9~eV in Fe, Ni and Cu,
respectively. 
The bare Coulomb interaction
$\langle \phi_d \phi_d |v|\phi_d \phi_d \rangle$ 
is actually $23.7$~eV, $25.9$~eV, and $27.4$~eV in Fe, Ni and Cu. 
This values become larger with increasing the number of 3$d$
occupation. 
Then the correlation term
$\langle \phi_d \phi_d |W^c(\omega =0)|\phi_d \phi_d \rangle$ 
is $-22.3$~eV, $-24.7$~eV and $-23.5$~eV in Fe, Ni and Cu.
Therefore, the correlation effects and the static screening 
are quite important in transition metals.
The screened correlation of Ni is largest and this is consistent with
the smallest $Z$ in these systems.
We should mention that the $d$-$d$ Coulomb interaction 
$\langle \phi_d \phi_d |W(0)|\phi_d \phi_d \rangle$ 
is different from the Hubbard $U$ evaluated from 
the constrained LSDA, which includes only screening by on-site $d$
electrons.  
The term  
$\langle \phi_d \phi_d |W(0)|\phi_d \phi_d \rangle$  
includes the screening effects by both on-site and 
off-site electrons.~\cite{Aryasetiawan-W}

% Figure ----------------------------------------------------------

In the transition metals, the Hubbard $U$ parameter is overestimated 
within the constrained LSDA, for example $U\approx6$~eV for Fe,
due to incomplete metallic screening in the LSDA.~\cite{U-metal}
We should mention that
our value of Ni with offset method is smaller than the previous
estimate
($\langle \phi_d \phi_d |W(0)|\phi_d \phi_d \rangle$
=2.2~eV).~\cite{Aryasetiawan-W}
The discrepancy may be caused by the absence of the present offset
method since  
we also obtained the value without the offset method similarly to be
previous one.

%%%%%%%%%%%%% Section4  %%%%%%%%%%%%%%%%%%%%%%%%%%%%%%%%%%%%%%%%
\section{Summary} \label{sec4:summary}

In this paper the $GW$ approximation is applied to ferromagnetic
transition metals Fe and Ni, and paramagnetic Cu.
We showed that the occupied 3$d$ band width of
transition metal is improved within our $GW$ calculation.

We also investigated the effects of dynamical screening by comparison
between the GWA and the static COHSEX approximation and showed the
crucial role of the dynamical correlation for band width.

The self-energy is discussed systematically.
The renormalization factor $Z$ showed that the interaction between
$d$-electrons in Ni is larger than that in Fe, and this is consistent
with the interaction strength from Auger spectra.

%%%%%%%%%%%%% Acknowledgements  %%%%%%%%%%%%%%%%%%%%%%%%%%%%%%%%
\section*{Acknowledgments}

We would like to thank to F. Aryasetiawan for fruitful discussions and
suggestions. 
We also thank to T. Kotani for useful suggestion about the offset
method. 
This work is supported by Grant-in-Aid for COE Research
``Spin-Charge-Photon'', and Grant-in-Aid from the Japan Ministry of
Education, Science, Sports and Culture. 
Part of the present calculation has been done by use of the facilities
of the Supercomputer Center, Institute for Solid State Physics,
University of Tokyo.

%%%%%%%%%%%%% Bibliography  %%%%%%%%%%%%%%%%%%%%%%%%%%%%%%%%%%%%

\end{document}